\documentclass[12pt]{spieman}  % 12pt font required by SPIE;
\usepackage{amsmath}
\usepackage{graphicx}
\usepackage{setspace}
\usepackage{tocloft}
\usepackage{epstopdf}
\usepackage{cite}
\usepackage{rotating}
\usepackage{multirow}
\usepackage[T1]{fontenc}
\usepackage{nicefrac}
\title{Scalable Platform for Adaptive optics Real-time Control (SPARC) Part 1: Concept, Architecture and Validation}

\author[a,*]{Avinash Surendran}
\author[b]{Mahesh P. Burse}
\author[b]{A. N. Ramaprakash}
\author[b]{Jyotirmay Paul}
\author[b]{Hillol K. Das}
\author[a]{Padmakar S. Parihar}
\affil[a]{Indian Institute of Astrophysics, Koramangala 2nd Block, Bangalore - 560034, India}
\affil[b]{Inter-University Centre for Astronomy and Astrophysics, No. 4, Ganeshkhind, Pune University Campus, Pune - 411007, India}

\cftpagenumbersoff{figure}
\cftpagenumbersoff{table} 
\begin{document} 
\maketitle

\begin{abstract}
We demonstrate a novel architecture for Adaptive Optics (AO) control based on FPGAs (Field Programmable Gate Arrays), making active use of their configurable parallel processing capability. SPARC's unique capabilities are demonstrated through an implementation on an off-the-shelf inexpensive Xilinx VC-709 development board. The architecture makes SPARC a generic and powerful Real-time Control (RTC) kernel for a broad spectrum of AO scenarios. SPARC is scalable across different numbers of subapertures and pixels per subaperture. The overall concept, objectives, architecture, validation and results from simulation as well as hardware tests are presented here. For Shack-Hartmann wavefront sensors, the total AO reconstruction time ranges from a median of 39.4$\mu$s ($11\times11$ subapertures) to 1.283 ms ($50\times50$ subapertures) on the development board. For large wavefront sensors, the latency is dominated by access time ($\sim$1 ms) of the standard DDR memory available on the board. This paper is divided into two parts. Part 1 is targeted at astronomers interested in the capability of the current hardware. Part 2 explains the FPGA implementation of the wavefront processing unit, the reconstruction algorithm and the hardware interfaces of the platform. Part 2 mainly targets the embedded developers interested in the hardware implementation of SPARC.
\end{abstract}


{\noindent \footnotesize*}Avinash Surendran,  \linkable{asurendran89@gmail.com}

\begin{spacing}{2}  

\section{Introduction}
\label{sect:intro}
Adaptive Optics (AO) is an indispensable tool in the frontier of astrophysics for the era of large telescopes of aperture diameter greater than 30 m, which can provide high-resolution spectroscopy and direct imaging of exoplanets. For example, the Narrow Field Infrared Adaptive Optics System (NFIRAOS)\cite{herriot2014} will be the first adaptive optics system deployed on the Thirty Meter Telescope (TMT) and will provide 12 times the spatial resolution (at near-infrared wavelengths) compared to that of the Hubble Space Telescope (HST). The Multi-Conjugate Adaptive Optics (MCAO) system on the European Extremely Large Telescope (E-ELT)\cite{diolati2008} will provide 16 times the spatial resolution compared to HST. AO is moving on from an optional back-end instrument to a full-time first light instrument on the next generation of large telescopes. Advances in the field of ground-layer AO\cite{tokovinin2004}, low noise sensors and better algorithms for wavefront sensing have opened up the possibility of ground layer AO using Natural Guide Stars (NGS) in the Giant Magellan Telescope (GMT)\cite{marcos2014}. The AO system on the GMT can provide a wide field correction over an 8" field-of-view (FOV) and has the capability for providing high contrast AO (80\% strehl ratio at a wavelength of 1.65 $\mu$m) for select targets as well\cite{hinz2010}. The WM Keck telescope was one of the first 10 m class telescopes to have an AO system (NGS in 1999 and LGS in 2004)\cite{chin2014} and have undergone repeated upgrades to provide a strehl ratio of upto 80\% at a wavelength of 1.65 $\mu$m\cite{wizinowich2008}. On the other end of the spectrum, small telescopes (of aperture diameter less than 5 m) can be used for long-term monitoring, has rapid responsivity to observe transient events and are available to a larger number of astronomers. An easily implementable and affordable AO system on a small telescope will expand its scientific capabilities by increasing the resolution and sensitivity of the same. The performance of AO on small telescopes have been demonstrated with a variety of implementations starting from the first Laser-Guide Star (LGS) AO system at the Lick Observatory\cite{max1997} to the more technologically advanced Robo-AO at the Palomar\cite{baranec2012} and Kitt-Peak telescopes\cite{jensen2017}. SPARC is a plug-and-play scalable generic AO kernel that will be powerful enough to cater to the computational requirements of large telescopes and is simple enough to be integrated with small telescopes. It is intended to make the integration of AO easier and more affordable for the entire spectrum of telescope sizes.

The landscape of AO, after a long period of evolution, is now prime for consolidation and standardization. Most of the deployed modern AO systems use Shack-Hartmann (SH) wavefront sensors\cite{wizinowich2006,rousset2000,dekany2013,rigaut2013,beuzit2008,gavel2002lick} (WFS) while a few use pyramid WFS\cite{morzinski2014} and curvature sensors\cite{rigaut1998,minowa2010}. The wavefront sensing is followed by the computation of a correction vector through AO reconstruction. Most of the currently deployed AO systems use matrix-vector-multiplication (MVM) in some form or the other. A multi conjugate adaptive optics (MCAO) system uses tomography and deformable mirror (DM) fitting to create the control matrix, which is only required to be created at a slower rate compared to the real-time reconstruction of actuator values from WFS inputs through MVM\cite{ellerbroek2013,wang2013}. The majority of the deployed and planned AO systems use a limited variety of WFS techniques, and MVM is predominantly used for extraction of the correction vector from the sensed variables.

We describe SPARC in two publications. Part 1 (this paper) explains the primary motive of the work, the hardware requirements, architecture, method of validation and the results from validation. This part is intended for astronomers and the user community who are primarily interested in understanding the capabilities and performance results of SPARC. Part 2\cite{SPARC2} describes in depth, the implementation of SPARC on an off-the-shelf Field Programmable Gate Array (FPGA) development board. Part 2\cite{SPARC2} contains the details of how scalability and adaptability has been built into SPARC in the memory interface and in the reconstruction matrix. It also explains the different techniques of AO reconstruction with their own levels of computational power required per degrees of freedom. Part 2\cite{SPARC2} is aimed at embedded system developers interested in the design of a scalable AO real-time controller on an FPGA.

Section~\ref{sect:oands} of this paper explains the objectives and the levels of scalability that is implemented on SPARC. Section~\ref{sec:cs} describes the SPARC system's state machine implementation and the details of the off-the-shelf hardware which we have used to test the platform. Details of the atmospheric turbulence simulation to check the limits of scalability and the interface with the iRobo-AO laboratory platform is explained in Section~\ref{sec:val}. Section~\ref{sec:res} gives the results obtained from the validation tests explained in Section~\ref{sec:val}. A summary and future prospects of the SPARC approach are presented in Section~\ref{sec:disc}.

\section{Objectives and Scalability}
\label{sect:oands}
Gavel \cite{gavel2002} had created an estimate of the computational requirements for achieving a Strehl ratio of 0.5 at a wavelength of 1$\mu$m for a telescope with a primary aperture diameter of 30 m. The fitting error in the error budget can be achieved by an AO system with 10,000 degrees of freedom (DOF). A more recent study of the AO requirement for E-ELT\cite{sevin2014} confirms a similar DOF target for E-ELT. The broader interface and computational requirements, which are based on this error budget, is given in Surendran et al\cite{asurendran2015}. The size of a control matrix for a single conjugate AO system with 10,000 DOF is close to 400 MB (if we assume 2 bytes for every element in the matrix). For an AO reconstruction time of 1 ms, a memory bandwidth of 400 GB/s and 400 GFlops of computing performance would be required. While the computing performance can easily be achieved by off-the-shelf GPUs and FPGAs, the memory bandwidth is still a challenge with the current technology. The Nvidia Tesla K10\cite{nvidia2014} is one of the most powerful GPUs available today and can provide a memory bandwidth of 320 GB/s.

A generic, scalable platform should be able to adapt to the requirements of large-scale AO and to the faster memory modules of the future. The primary motive of SPARC is to create a {plug-and-play} implementation of a scalable adaptive optics real-time controller. FPGAs are ideally suited for such a system because of its inherent flexibility in adapting the level of parallelization based on requirement. This translates to a standalone AO real-time controller with flexible WFS and DM interfacing options and is scalable and adaptable to a range of AO scenarios as follows:
\begin{enumerate}
\item Number of pixels and pixels per subaperture: One of the largest contributors to the error budget of an AO system is the fitting error caused by the limitations in the spatial sampling of the wavefront. A unique feature of the SPARC implementation is that the logic resources of the FPGA needed for AO computation (including pixel acquisition, slope computation, AO reconstruction and actuator output) are independent of the number of subapertures. The slope computation takes a relatively small fraction of the logic resources and hence does not contribute much to the FPGA resource usage. The platform is also compatible with rectangular or cropped subaperture configurations.
\item Memory bandwidth: An adaptable First-in-first-out (FIFO) interface is used in SPARC which provides a lot of flexibility in interfacing with dual data rate (DDR) type memory modules with a wide range of frequencies and datawidths.
\item Portability across FPGA families: SPARC is designed with compatibility in mind, where the core algorithm is programmed with native VHDL, which is compatible with the FPGAs manufactured by most companies. About 25\% of the program is FPGA specific but this is restricted to the operation of external interfaces. This enables the design to be implemented on slower/faster FPGAs having less/more internal memory and logic resources, which can interface with slower/faster external memory modules.
\end{enumerate}

SPARC is designed to function without the help of a host-PC for the AO interfaces, which will help in the miniaturization of the computational hardware. This is especially advantageous for small AO systems on a budget, if the control system is able to plug-and-play into the optical hardware. The {plug-and-play} implementation can be expanded to other applications of AO like vision science and microscopy. An FPGA is used as the {computational} hardware for the following reasons:
\begin{enumerate}
 \item Flexibility in parallel processing: The number of parallel processes in a GPU or CPU is limited by the number of processor cores available on hardware and a single clock driving the chip, whereas the FPGA is only limited by the raw logic available inside the chip. An FPGA allows for the creation of a custom number of parallel processing units (limited only by the total logic resources available) and to create different digital clocks whose periods can be individually customized. Matrix multiplications can be optimized with the available memory bandwidth with this capability.
 \item Predictable latency: The delay between the time at which input signals from the wavefront sensor arrives to the time at which the deformable mirror correction signals are updated, directly affects the performance of an AO system. Unlike CPUs and GPUs, the computational delay of an FPGA design (excluding the delay contributed by external interfaces) can be predicted to within a few nanoseconds.
\end{enumerate}

\section{Control Scheme}
\label{sec:cs}
\subsection{Hardware}
The SPARC platform is implemented on a commercial off-the-shelf solution which could demonstrate the wide range of scalability and is powerful enough to execute the AO control loop for a large telescope. The Xilinx VC-709 development board has a Virtex-7 XC7VX690T chip which hosts 3,600 Digital Signal Processors (DSPs) and facilitates a large number of parallel multiply-and-accumulate (MAC) operations for a fast matrix multiplication. The two modules of DDR3 (DDR version 3) memory on the board can provide a combined memory bandwidth of 25.6 GB/s, which is important for the fast retrieval of the reconstruction matrix for AO reconstruction. The board also features a Peripheral Control Interconnect express (PCIe) version 3 which facilitates high speed communication with a host PC.

\subsection{State machine implementation of the control loop}

The FPGA design is divided into state machines which are a combination of digital logic gates for computation (combinatorial logic) and decision making algorithms (sequential logic). The sequential logic in each state machine is driven by periodic clocks whose frequency is determined by the maximum time taken by the logic gates to complete a computation within the state machine. The design consists of three state machines, which are driven by different clock frequencies owing to the different amount of time taken for computation by the logic gates within each state machine.

The functioning of SPARC is like any other AO control loop which accepts WFS pixels as the input and generates phase (or actuator) values as the output. The difference is the scalability which is incorporated at each level. Fig.~\ref{fig:sm} shows SPARC represented as a system of interconnected state machines. The state machine shows the flow of data for a single row of subapertures coming from the WFS. The three state machines are described below:

\begin{figure}
\centering
\includegraphics[width=0.9\textwidth]{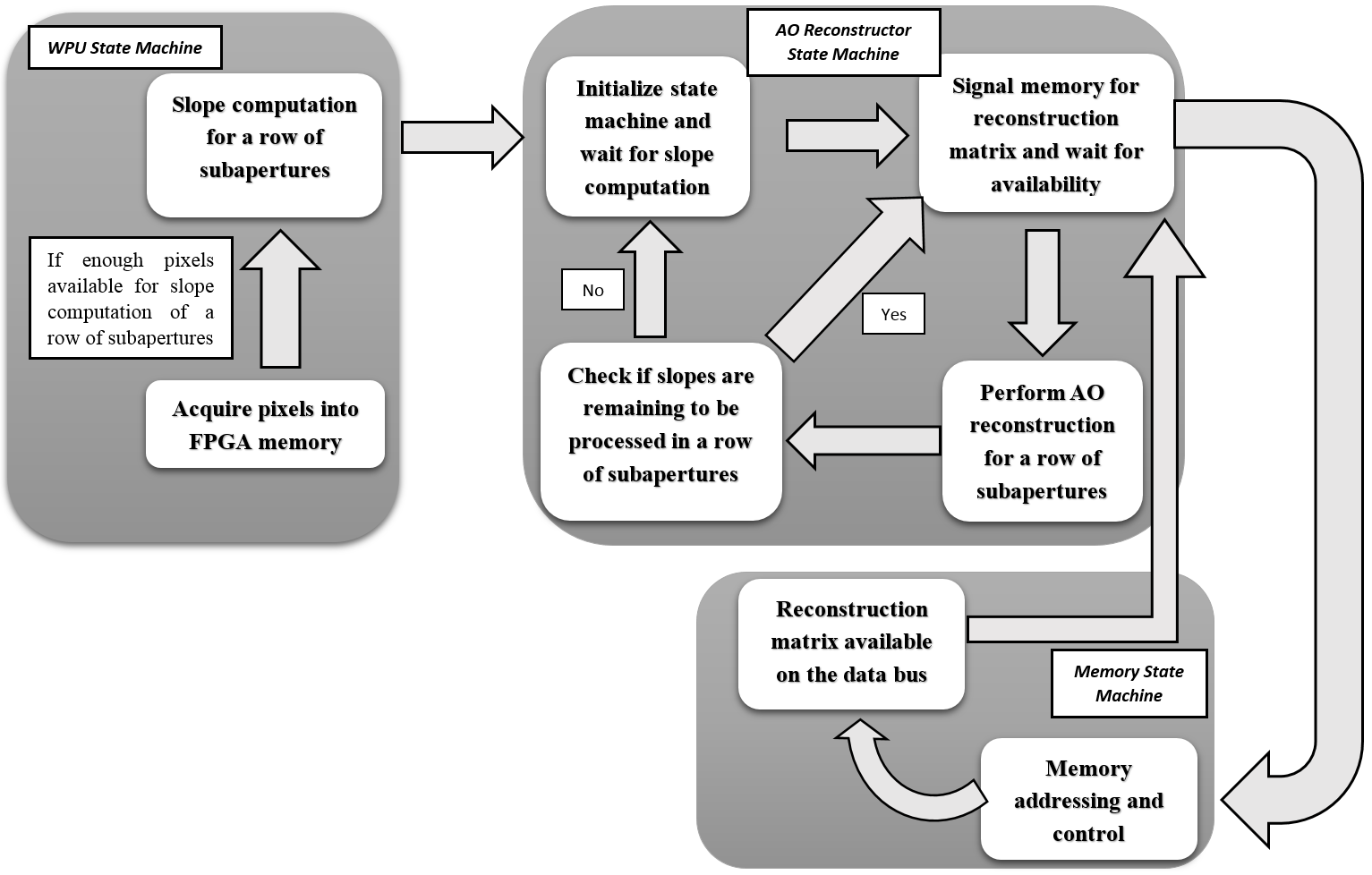}
\caption{State Machine representation of SPARC, showing a simplified schematic of how data flows from the WFS to the matrix multiplication corresponding to a single row of subapertures.}
\label{fig:sm}
\end{figure}

\begin{enumerate}
 \item Wavefront Processing Unit (WPU): The WPU consists of a flexible interface for acquiring pixels from different CCD interfaces, and a scalable slope computer. This particular WPU is designed to implement a Center of Gravity (CoG) slope computation for an SH sensor, but the algorithm for slope computation is modular and can be modified if required. Every CCD interface has its own set of rules on how the pixels are sent to the processing hardware. So, a flexible interface was provided, which can be connected to the custom WFS hardware, or to a host PC which can acquire the pixels before sending them to the FPGA. Each pixel value is 16 bits in the current hardware, but this can be changed if required. The WPU is pipelined to allow for simultaneous computation of slopes without interrupting the pixel acquisition into the FPGA. When the pixels corresponding to a single row of subapertures is available, the slope computation begins. {The mapping of pixels to subapertures is done through a flexible Block RAM (BRAM) addressing algorithm (according to number of subapertures and pixels per subaperture). The pixels of the same subaperture are written into BRAM in a way that all of those pixels can be simultaneously accessed in a single clock cycle for slope computation. Currently, SPARC supports any square array of pixels in a subaperture. More details on the BRAM addressing is specified in Surendran et al\cite{asurendran2015} and Part 2\cite{SPARC2} of the paper.} The speed of slope computation is determined by the amount of logic resources on the FPGA. A combination of dynamic internal memory allocation, and the use of different clocks for pixel acquisition and slope computation has resulted in a WPU which is fully scalable with the number of subapertures and pixels per subaperture. The logic resource usage in the FPGA only depends on the pixels per subaperture and the speed of slope computation (which can be set by the user), and is independent of the number of subapertures. {As of now, the implementation of SPARC on the VC-709 development board requires a host PC with a PCIe interface, to send and receive data to the outside world (including the WFS and the DM).}
 \item AO Reconstructor: The AO Reconstructor analyzes the reconstruction matrix, decomposes it into sub-matrices depending on the number of subapertures in the WFS and performs MVM for each row of subapertures at a time. We are currently using Fried geometry in our platform, but any geometry can be used to create the reconstruction matrix. If the number of subapertures along a row is $n$ (and the total number of subapertures $n^2$), the size of the reconstruction matrix would be $(n+1)^2\times2n^2$ (for Fried geometry). The x-slopes and y-slopes from each row of subapertures would correspond to $2n$ slopes, and they would need to be multiplied with two sub-matrices of size $(n+1)^2 \times n$ taken from different parts of the original reconstruction matrix. As shown in Fig.~\ref{fig:sm}, the AO reconstructor communicates with the memory state machine to extract the correct sub-matrix (corresponding to the available slopes) from the external memory, and performs the required multiplication. The speed for this is limited only by how fast the sub-matrix can be read out from the external memory. The details on how the memory speed adaptability and scalability is implemented in the AO reconstructor, is discussed in part 2\cite{SPARC2} of the paper.
 \item Memory state machine: As mentioned in Section~\ref{sect:intro}, the speed of current external memory modules available in the market is not enough to cater to the requirements of AO for large telescopes. Hence, the SPARC memory interface has been designed to be flexible to cater to different frequencies and datawidths of the external memory being interfaced. {The detailed description of the scalability of the memory interface is provided in Part 2\cite{SPARC2} of the paper.} The memory state machine is also responsible for acquiring the reconstruction matrix from a host PC and writing it into the external memory connected to the FPGA.
\end{enumerate}

\section{Validation}
\label{sec:val}

Validation of SPARC is performed through two methods:
\begin{enumerate}
 \item Hardware-in-the-loop simulation: In this method, atmospheric turbulence is simulated for different telescope aperture sizes and the performance of SPARC tested for a range of subaperture sizes. This simulation measures the AO reconstruction time and verifies the phase outputs generated by SPARC.
 \item iRobo-AO interface: In this test, SPARC is interfaced with actual AO hardware. Reliability testing for a large number of frames is conducted by using SPARC as the real-time AO kernel in a real AO system called iRobo-AO (see below). This was not possible with the hardware in the loop test due to the long duration of time that it took for the host PC to process each frame.
\end{enumerate}

\subsection{Hardware-in-the-loop simulation}
The AO simulator is based on the fast-fourier transform based phase screens for large telescopes designed by Sedmak\cite{sedmak2004}, and is created with MATLAB. A master phase screen is created initially, from which are generated the continuous individual phase screens which travel across the primary aperture (frozen flow). The frozen flow is generated based on the number of frames, wind speed and frames per second. The input simulated wavefront has not been corrected for tip-tilt. The generated frames from the phase screen are used to create the WFS pixel inputs. {The WFS pixels are generated from the phase screen assuming Fried geometry, as described by Herrmann\cite{herrmann1980}.} The PCIe interface of the VC-709 development board is used for sending simulated pixel inputs from the host PC to SPARC and to retrieve the phase outputs generated by SPARC. At present, the simulator accounts for fitting error and time delay error. Other error sources can be easily incorporated in future revisions. 

\begin{figure}
\centering
\includegraphics[width=1\textwidth]{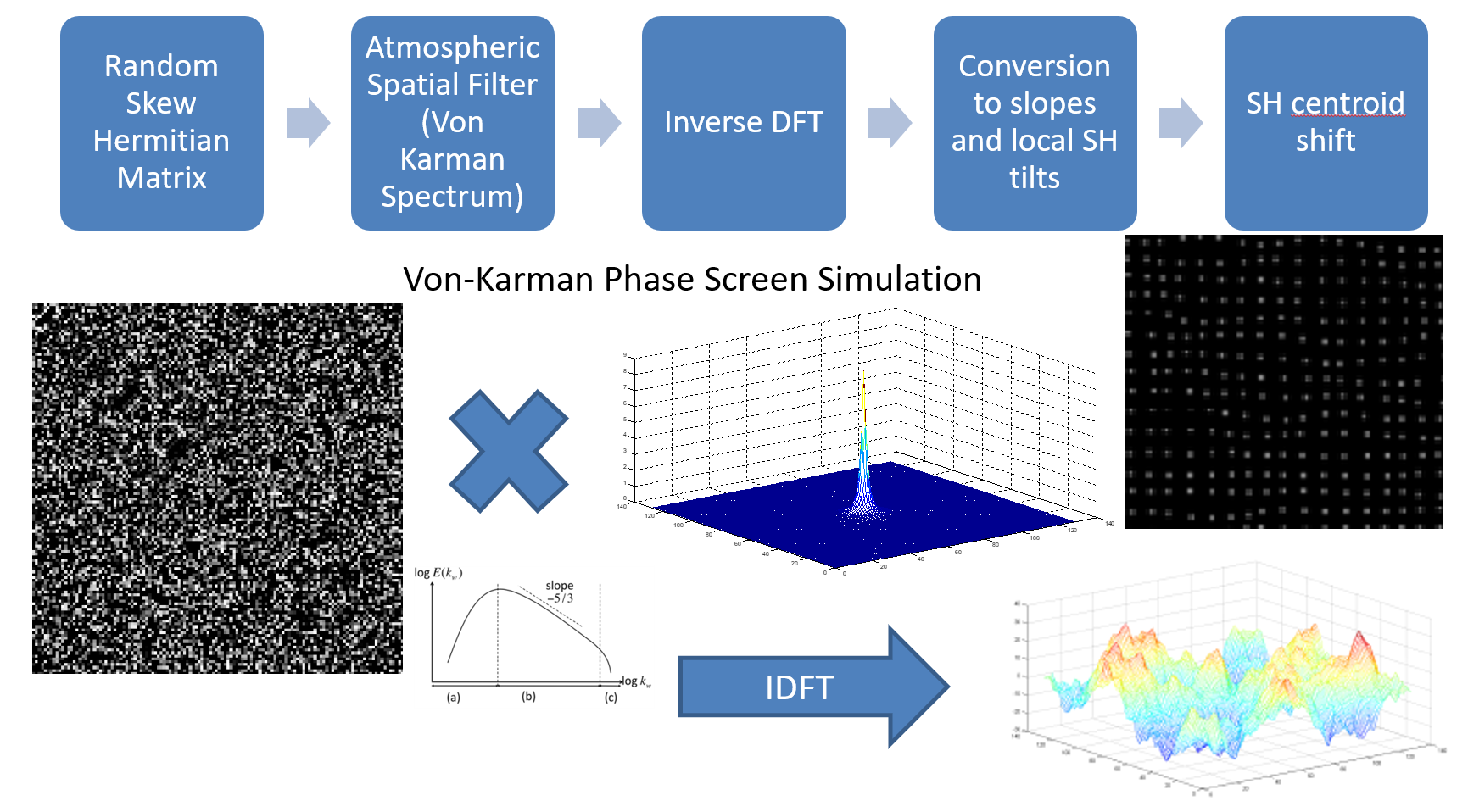}
\caption{Schematic of phase screen generation, based on the method outlined by Sedmak\cite{sedmak2004}. The inverse digital Fourier transform (DFT) of the product of a random skew-hermitian matrix (with a mean of 0 and variance of 1) with the Von-Karman power spectrum produces the required phase screen for any aperture diameter.}
\label{fig:pcie}
\end{figure}

\begin{figure}
\centering
\includegraphics[width=0.7\textwidth]{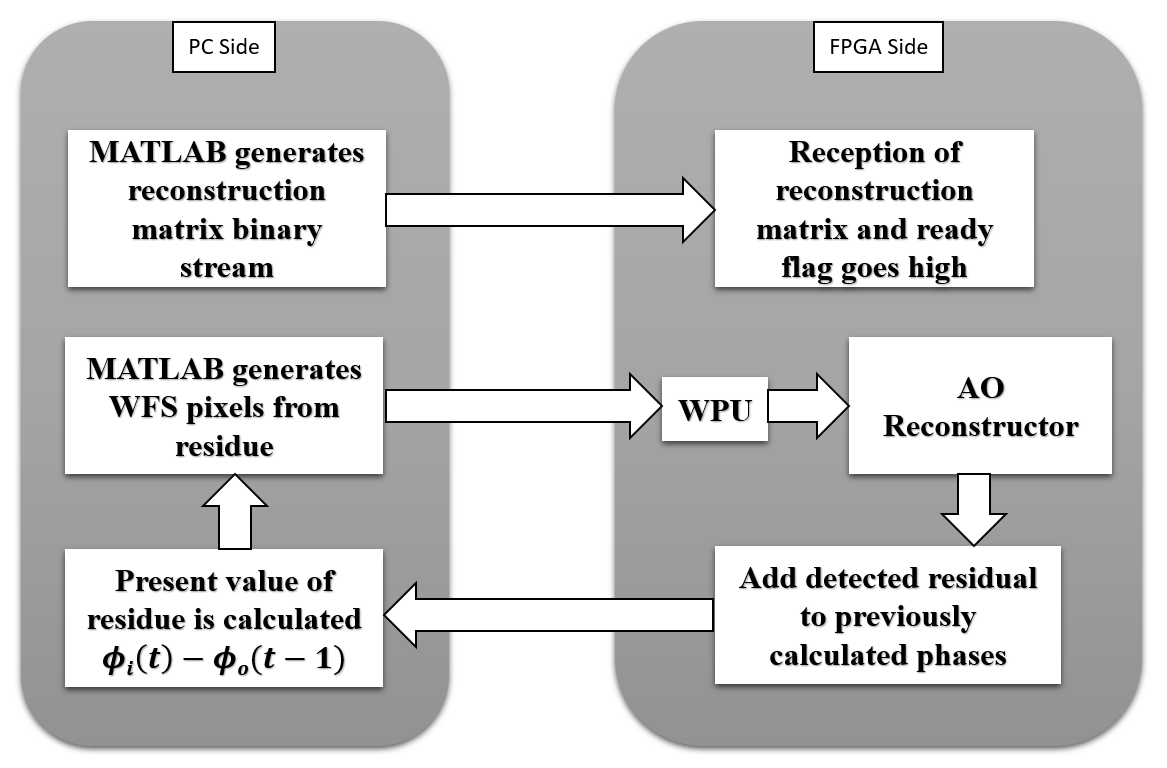}
\caption{Hardware-in-loop simulation for SPARC, which shows the PC generating the simulated pixels of a WFS. After SPARC generates the phase outputs at the FPGA side and sends it to the PC, the same simulator generates the residual wavefront and the next frame of WFS pixels to be sent to SPARC.}
\label{fig:pcie}
\end{figure}

Fig.~\ref{fig:pcie} shows the dataflow which is used for testing the platform. After power up of the FPGA development board, the reconstruction matrix is sent to the DDR3 memory through the PCIe interface. In {Fig~\ref{fig:pcie}}, $\phi_i(t)$ represents the individual phase screen generated by the MATLAB program and $\phi_o(t-1)$ represents the reconstructed wavefront derived by the FPGA from the previous WFS frame. At the start of the AO loop, $\phi_o(t-1)$ is zero. The number of frames in our simulations vary from 1000-2000, and can be set from the PC.

The benchmarking carefully excludes the performance of the PC interfaces involved, and only captures the performance of the AO real-time kernel implemented on the FPGA. The test program simulates the phase screen for 1000 frames per second (fps) for 1 - 2 seconds. But the computation of the residual phase for each WFS frame by the MATLAB AO simulator in the PC takes a few seconds to a few minutes (depending on the size of the WFS frame). The platform was tested for subaperture values ranging from $11\times11$ to $50\times50$ subapertures. The pixels per subaperture is set to either $2\times2$ or $4\times4$ pixels.

\subsection{iRobo-AO interface}

\begin{figure}
\centering
\includegraphics[width=0.9\textwidth]{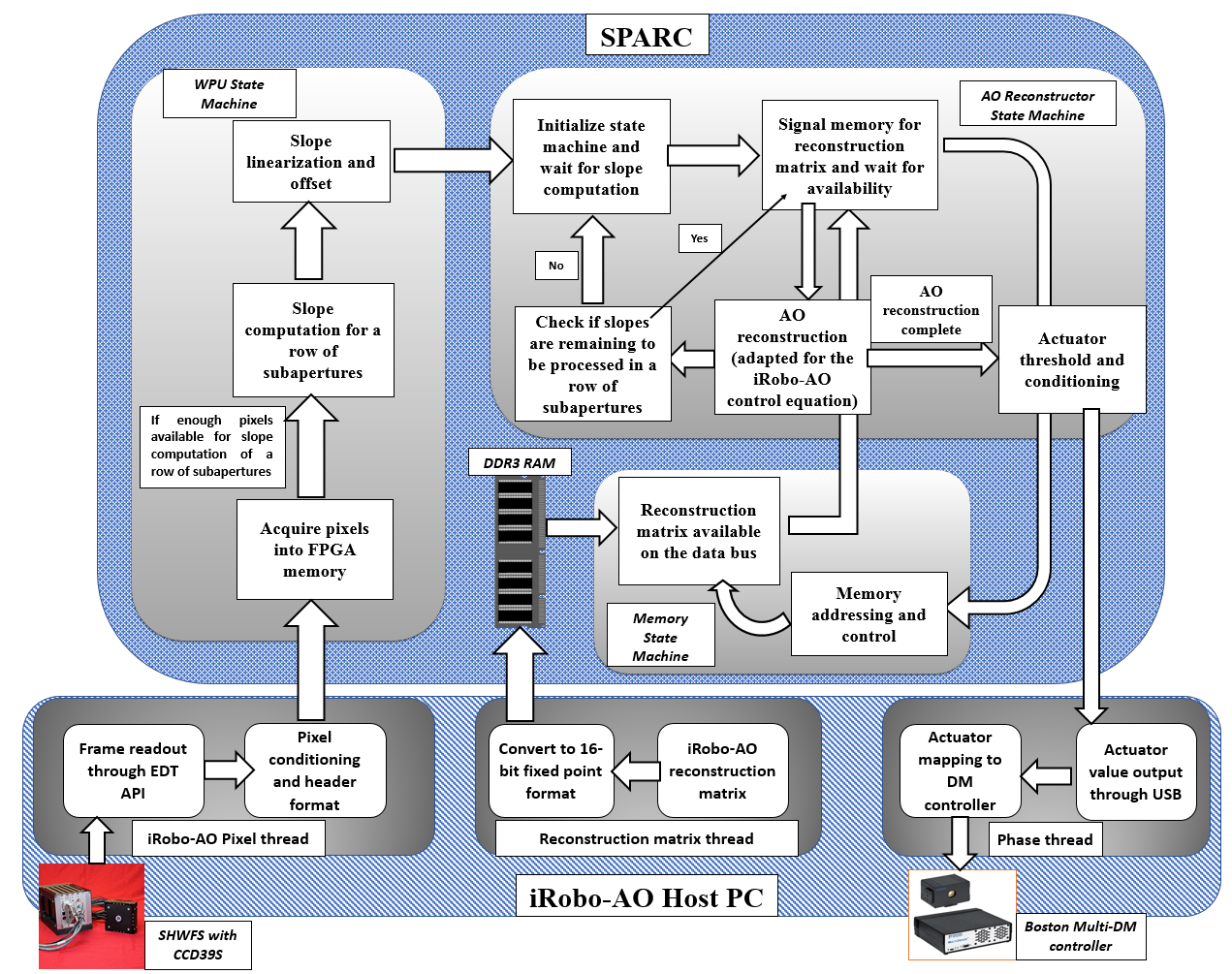}
\caption{Schematic of the interface with the iRobo-AO laboratory platform. The PC side consists of a multithreaded software which processes and transmits WFS pixels and actuator values from the WFS to SPARC and from SPARC to the DM respectively. The schematic of how SPARC was modified for working with iRobo-AO is shown in the inset.}
\label{fig:iRobo-AO}
\end{figure}

The iRobo-AO system is a robotic AO system due to be installed at the Inter-University Centre for Astronomy and Astrophysics (IUCAA) Girawali Observatory shortly. The first Robo-AO was installed at the Palomar P60 telescope\cite{baranec2012} as well as the Kitt Peak 2m telescope\cite{jensen2017}, and has been used for several years now to take thousands of observations at high resolution. {The Robo-AO WFS consists of 97 illuminated subapertures and 120 DM actuators. The iRobo-AO system is virtually of the same design as the Robo-AO system, except for the optical interface which is tailored to the Girawali Observatory.}The main difference in the the iRobo-AO interface from the hardware-in-the-loop simulation is the replacement of the simulated wavefront and computed feedback, with real-time optical feedback of the corrected wavefront from DM and the WFS.

Fig.~\ref{fig:iRobo-AO} shows how the state machines of SPARC (given in Fig.~\ref{fig:sm}) interact with the host PC and the interfaces of iRobo-AO. Before the initialization of the AO loop, the iRobo-AO reconstruction matrix is to be sent to the external DDR3 memory on the FPGA development board. The size of the Robo-AO reconstruction matrix is $120\times194$. The size is derived from 120 illuminated DM actuators and the 194 wavefront slopes (two slopes each from 97 illuminated subapertures). SPARC accepts rectangular or square subaperture shapes, and the reconstruction matrix has to be padded with zeros for the non-illuminated subapertures to create a reconstruction matrix of size $144\times242$ ($12\times12$ actuators mapped from two slopes each for $11\times11$ subapertures). The matrix is converted into 16-bit binary fixed point format at the host PC, and sent to the FPGA through the PCIe interface.

The reconstruction matrix transmission is a one time operation, and is followed by the iRobo-AO pixel thread (at the host PC) which acquires the first WFS frame. The SHWFS consists of $11\times11$ subapertures which image SH spots on to a $80\times80$ pixel CCD39S frame-transfer charge coupled device (CCD) and is connected to the host PC via a cameralink interface. After dark subtraction and $3\times3$ binning of pixels, the $22\times22$ binned pixel area corresponding to the square illuminated section of the CCD is cropped and relevant headers are added. The pixel data stream converts the pixels into 16-bit binary data and sends it to the FPGA via the pixel data stream through the PCIe interface. The WPU state machine starts the conversion of WFS pixels into slopes after the pixels corresponding to every row of subapertures become available to the WPU. For iRobo-AO, SPARC is adapted to perform slope linearization and offset according to the corresponding lookup tables, which are fed at the time of programming the FPGA. Slope linearization and offset was required to compensate for the non-linearity in centroid estimation (inherent in SH sensors) and to offset the centroids for when the DM is flat, respectively. The slope computation is followed by the AO reconstruction of the corresponding row of subapertures. The AO control equation has also been adapted with gain and leak parameters, to replicate the iRobo-AO control loop equation. The interface with iRobo-AO was also a test of the flexibility of SPARC to work as a simulator as well as with a real AO system. The adaptations to SPARC were modular, and similar features can be seamlessly added to it without the pain of overhauling the entire platform.

After AO reconstruction for a single frame of WFS is completed, SPARC sends 144 actuator values through the PCIe interface to the actuator data stream. The actuator data stream is always active, waiting for actuator values from the FPGA and functions independently from the other subroutines on the host PC. The actuator values are subjected to optional DM safety threshold values, and are then mapped to the physical pins of the DM controller. The four corner actuator values are discarded and 140 actuator values are sent to the DM controller.

\section{Results}
\label{sec:res}

\subsection{Results from hardware-in-the-loop simulation}

The validation of the AO correction is done by comparing the theoretical RMS WFE\cite{roddier1999} due to a combination of fitting error and time delay error, with the actual RMS WFE produced by the hardware-in-the-loop simulation. {The AO system is assumed to adhere to a strict Fried geometry with the number of phase values generated being $(n+1)^2$, if the number of subapertures is $n^2$.} The results reported here are obtained with a Fried parameter ($r_0$) of 15 cm, a mean wind velocity ($\overline{v}$) of 5 m/s, an outer scale of turbulence ($L_0$) of 25 m {and a modelled loop frequency of 1 kHz}. The telescope aperture diameter is varied for a change in the number of subapertures to ensure that positions of the SH spots are well within the dynamic range of the simulated Shack-Hartmann wavefront sensor (SHWFS). 

\subsubsection{Performance gain}

\begin{figure}
\centering
\includegraphics[width=0.9\textwidth]{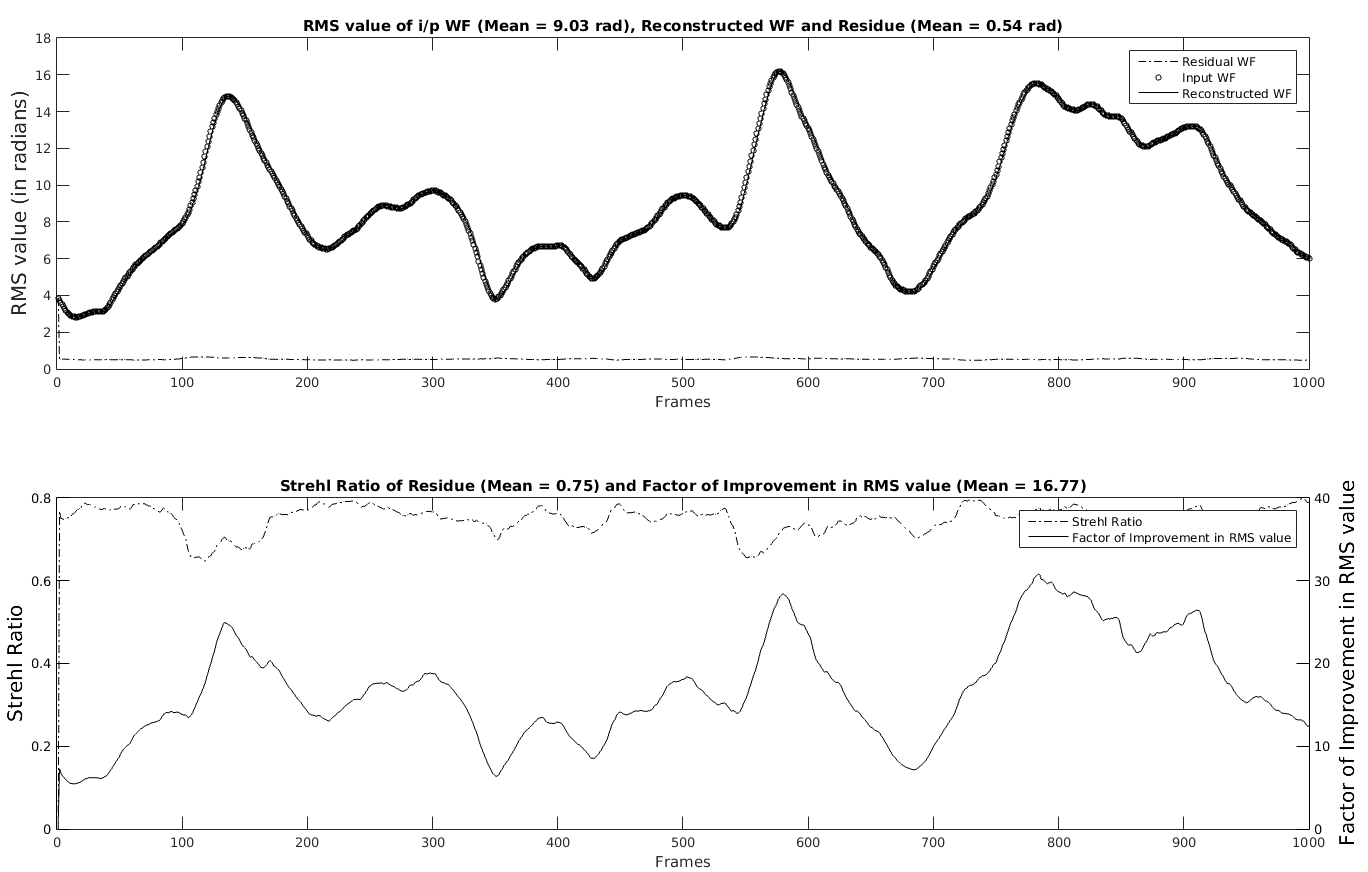}
\caption{Results from a $50\times50$ subaperture AO hardware-in-the-loop simulation for an aperture diameter of 8 m. a) The RMS WFE of the input wavefront, the reconstructed wavefront (from the FPGA) and the residual wavefront b) Strehl ratio and factor of improvement in the RMS WFE, over 1000 frames of the hardware-in-the-loop simulation.}
\label{fig:50x50}
\end{figure}

\begin{figure}
\centering
\includegraphics[width=0.9\textwidth]{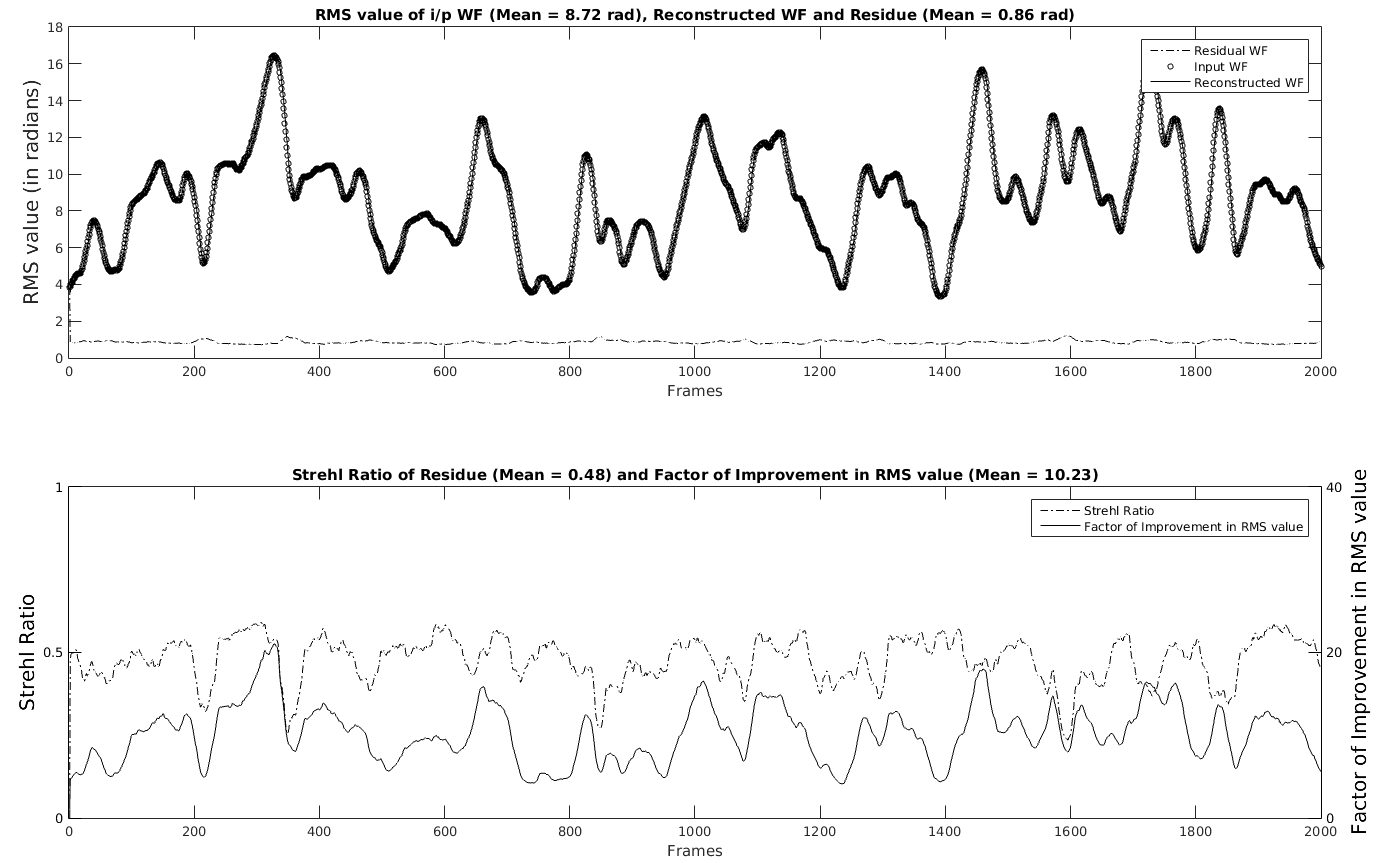}
\caption{Results from a $42\times42$ subaperture AO hardware-in-the-loop simulation for an aperture diameter of 8 m. a) The RMS WFE of the input wavefront, the reconstructed wavefront (from the FPGA) and the residual wavefront b) Strehl ratio and factor of improvement in the RMS WFE, over 2000 frames of the hardware-in-the-loop simulation.}
\label{fig:42x42}
\end{figure}

Fig.~\ref{fig:50x50} shows the results of the hardware-in-the-loop simulation with $50\times50$ subapertures for an aperture diameter of 8 m. Fig.~\ref{fig:42x42} shows the results of the hardware-in-the-loop simulation with $42\times42$ subapertures for the same aperture diameter of 8 m. The figures show the time variation of the instantaneous RMS values of the input wavefront, the reconstructed wavefront and the residual wavefront on the SHWFS, $\phi_i(t) - \phi_o(t - 1)$. The Strehl ratio (by Marechal approximation) and the instantaneous factor of improvement between the RMS values with and without AO correction, are also shown in the figures. The figures show the comparison of the change in the strehl ratio for different number of subapertures, simulated at the same aperture diameter of 8 m. The fitting error will be higher for a smaller number of subapertures due to a lower spatial sampling, and hence the mean strehl ratio is smaller for $42\times42$ subapertures compared to $50\times50$ subapertures.

{Table~\ref{tab:ao} shows the results of AO correction at different subaperture sizes. Fitting error for an AO system is given by the equation\cite{roddier1999},}
$$\sigma^2_{fit}=0.335\bigg(\frac{D}{r_0\times n^2}\bigg)^{\nicefrac{5}{3}}$$
{where $D$ is the aperture diameter, $r_0$ is the Fried parameter and $n^2$ is the number of independently controlled actuators in the AO system. Under the Fried geometry assumption, the number of actuators is one more than the number of subapertures along a row. The number of subapertures for each aperture diameter is chosen such that the ratio of $\nicefrac{D}{n^2}$ ranges between 10 - 20 cm for all observations. Time delay error is constant because of the constant modelled loop frequency that we used, hence keeping the combination of fitting and time delay error similar for different observations in Table~\ref{tab:ao}.}

\begin{table}[]
\centering
\caption{AO correction at different subaperture sizes and telescope aperture diameters. The input wavefront is simulated with a Fried parameter ($r_0$) of 15 cm, a mean wind velocity ($\overline{v}$) of 5 m/s and an outer scale of turbulence ($L_0$) of 25 m}
\label{tab:ao}
\begin{tabular}{lllllll}
\multirow{2}{*}{Subapertures} & \multirow{2}{*}{\begin{tabular}[c]{@{}l@{}}Aperture\\ Diameter\\ (in meters)\end{tabular}} & \multicolumn{3}{c}{Mean of the RMS values (in radians)}                                                                                                                                                          & \multirow{2}{*}{\begin{tabular}[c]{@{}l@{}}Strehl\\ Ratio\end{tabular}} & \multirow{2}{*}{\begin{tabular}[c]{@{}l@{}}Factor of\\ Improvement\\ in RMS WFE\end{tabular}} \\
                              &                                                                                            & \begin{tabular}[c]{@{}l@{}}Input\\ wavefront\end{tabular} & \begin{tabular}[c]{@{}l@{}}Residual\\ wavefront\end{tabular} & \begin{tabular}[c]{@{}l@{}}Fitting +\\ Time delay\\ error\end{tabular} &                                                                         &                                                                                               \\
$50\times50$                         & 8                                                                                          & 9.03                                                      & 0.54                                                         & 0.62                                                                   & 0.75                                                                    & 16.77                                                                                         \\
$42\times42$                         & 6                                                                                          & 5.10                                                      & 0.43                                                         & 0.57                                                                   & 0.83                                                                    & 12.07                                                                                         \\
$32\times32$                        & 5                                                                                          & 3.44                                                      & 0.33                                                         & 0.60                                                                   & 0.90                                                                    & 10.63                                                                                          \\
$21\times21$                         & 4                                                                                          & 2.18                                                      & 0.24                                                         & 0.70                                                                   & 0.94                                                                    & 9.25                                                                                          \\
$16\times16$                         & 3                                                                                          & 1.49                                                      & 0.27                                                         & 0.68                                                                   & 0.93                                                                    & 5.64                                                                                          \\
$11\times11$                         & 2                                                                                          & 0.99                                                      & 0.20                                                         & 0.65                                                                   & 0.96                                                                    & 4.92                                                                                         
\end{tabular}
\end{table}

\begin{table}[]
\centering
\caption{Comparison of the standard deviation of the RMS values of the input wavefront to the standard deviation of the RMS values of the residual wavefront}
\label{tab:sd}
\begin{tabular}{lllll}
Subapertures & \begin{tabular}[c]{@{}l@{}}Aperture \\ Diameter\\ (in metres)\end{tabular} & \begin{tabular}[c]{@{}l@{}}Standard deviation \\ of input WF ($S_i$)\end{tabular} & \begin{tabular}[c]{@{}l@{}}Standard deviation \\ of residual WF ($S_r$)\end{tabular} & \begin{tabular}[c]{@{}l@{}}Ratio \\ $\nicefrac{S_i}{S_r}$\end{tabular} \\
50x50        & 8                                                                          & 3.39                                                                             & 0.11                                                                                & 30.82                                                        \\
42x42        & 6                                                                          & 2.31                                                                             & 0.23                                                                                & 10.04                                                        \\
32x32        & 5                                                                          & 1.14                                                                             & 0.04                                                                                & 28.50                                                        \\
21x21        & 4                                                                          & 0.85                                                                             & 0.05                                                                                & 17.00                                                        \\
16x16        & 3                                                                          & 0.56                                                                             & 0.06                                                                                & 9.33                                                         \\
11x11        & 2                                                                          & 0.42                                                                             & 0.03                                                                                & 14.00                                                       
\end{tabular}
\end{table}

The mean of the RMS values (calculated as the ratio of the sum of the RMS values of all the wavefronts to the total number of WFS frames) of the uncorrected wavefront (calculated over the number of frames) is about 20-30\% of the theoretically estimated value\cite{roddier1999}, because of undersampling of the Von Karman spectrum at low frequencies. Since the number of samples in the frequency domain is the same as that of the phase screen in the time domain (equal to the number of actuators), the undersampling error is inversely proportional to the number of subapertures for which the phase screen is simulated. In the simulation, we have used uniform sampling without any sub-sampling below the lowest frequencies. {A lower RMS value for the residual WF is obtained for smaller subapertures, because the input WF gets more undersampled as the number of subapertures reduces.} Considering that, the mean of the RMS values of the residual wavefronts at different subaperture sizes are well within the estimated theoretical values of the error budget due to fitting error and time delay error. {Table~\ref{tab:sd} shows the comparison between the standard deviations of the RMS values of the uncorrected wavefront and the residual wavefront.}

\subsubsection{AO Reconstruction time}

\begin{figure}
\centering
\includegraphics[width=1\textwidth]{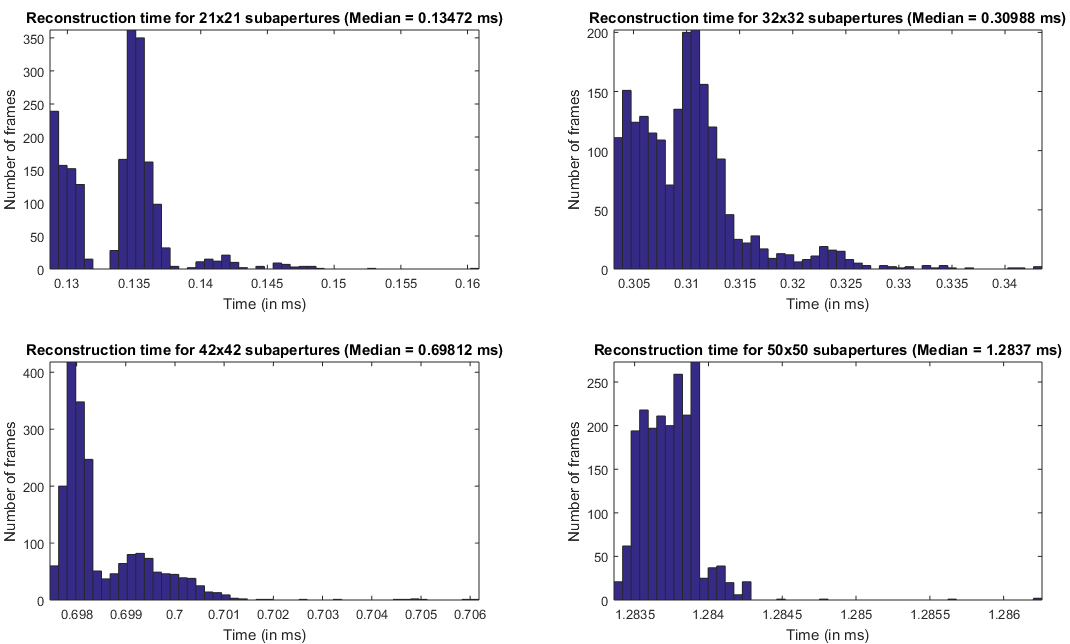}
\caption{Variability in AO Reconstruction time for $21\times21$, $32\times32$, $42\times42$ and $50\times50$ subaperture sizes. The results were generated from a hardware-in-the-loop simulation with 2000 WFS frames.}
\label{fig:runtime}
\end{figure}

The time taken for AO reconstruction of a WFS frame is calculated from when the first pixel arrives at the input of SPARC to when the last actuator value is sent out to the host PC. As explained in Section~\ref{sec:val}, care has been given to exclude the time taken for the WFS pixel and the DM actuator values to pass through the host PC interfaces. Fig.~\ref{fig:runtime} shows the histogram of the variability in the AO reconstruction time. The results in Fig.~\ref{fig:runtime} are obtained at $4\times4$ pixels per subaperture. The variation between the slowest and fastest frame is within a few tens of microseconds, and is solely caused by the unpredictable latency of the external DDR3 memory. All other processes except fetching the reconstruction matrix from the external memory are deterministic and repeatable across frames. For the largest number of subapertures ($50\times50$) that we simulated SPARC with, 1 ms out of a total AO reconstruction time of 1.283 ms was taken in the retrieval of the reconstruction matrix from the DDR3 memory. For all the subaperture values that we tested SPARC with, the time take for reconstruction matrix retrieval formed a very significant percentage of the total reconstruction time. A comparison between the mean reconstruction time and the time taken for fetching the reconstruction matrix from an ideal DDR3 memory without any unpredictable latencies is shown in Part 2\cite{SPARC2} of this paper.

\subsection{Results from iRobo-AO interface}
\begin{figure}
\centering
\includegraphics[width=1\textwidth]{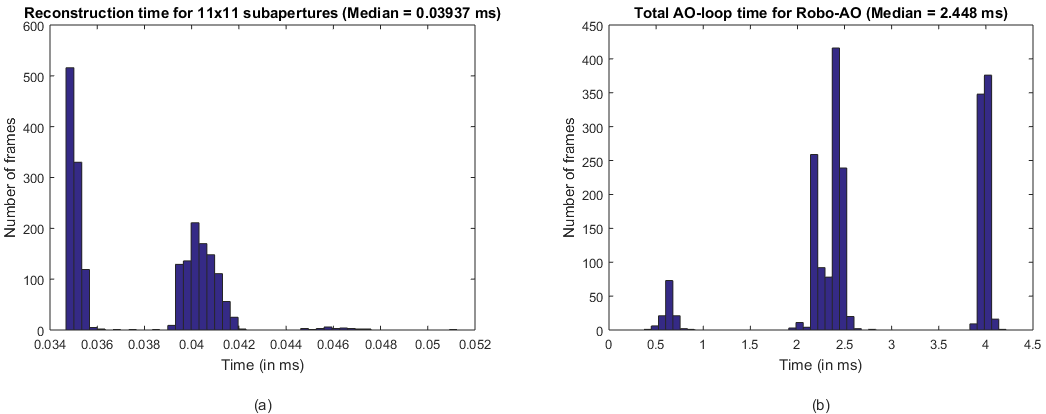}
\caption{Comparison of a) AO reconstruction time (excluding external interfaces) and b) total AO loop time (including cameralink and PCIe interface time) for the SPARC system interfaced with iRobo-AO.}
\label{fig:iroboao_runtime}
\end{figure}

The multithreaded operation of the pixel thread and the phase thread at the host PC minimized the latency between the WFS and the DM of iRobo-AO. Fig~\ref{fig:iroboao_runtime} shows the comparison between the total reconstruction time per frame and the total AO loop time for a set of 2,000 frames of the SPARC working with iRobo-AO. The median reconstruction time (inside the FPGA) without the time taken for the data to pass through the interfaces comes to less than 40 ns per frame. The AO loop time is much higher because of the low WFS frame rate of around 550-600 frames/second. The interfaces for the PCIe, the WFS and the DM are outside the scope of implementation of SPARC. They are part of the testing platform, and the performance of these interfaces were not benchmarked or improved upon. The same latencies would be present in a GPU implementation, apart from additional internal memory and processor-to-processor latencies\cite{altera2013}. We were using an off-the-shelf PCIe software, where the data sent through the interface had to be checked for integrity and resent if it was erroneous. This resulted in the three groups of AO loop times (in Fig~\ref{fig:iroboao_runtime}b) according to the number of times the data had to be resent through the interface to preserve data integrity.

\begin{figure}
\centering
\includegraphics[width=1\textwidth]{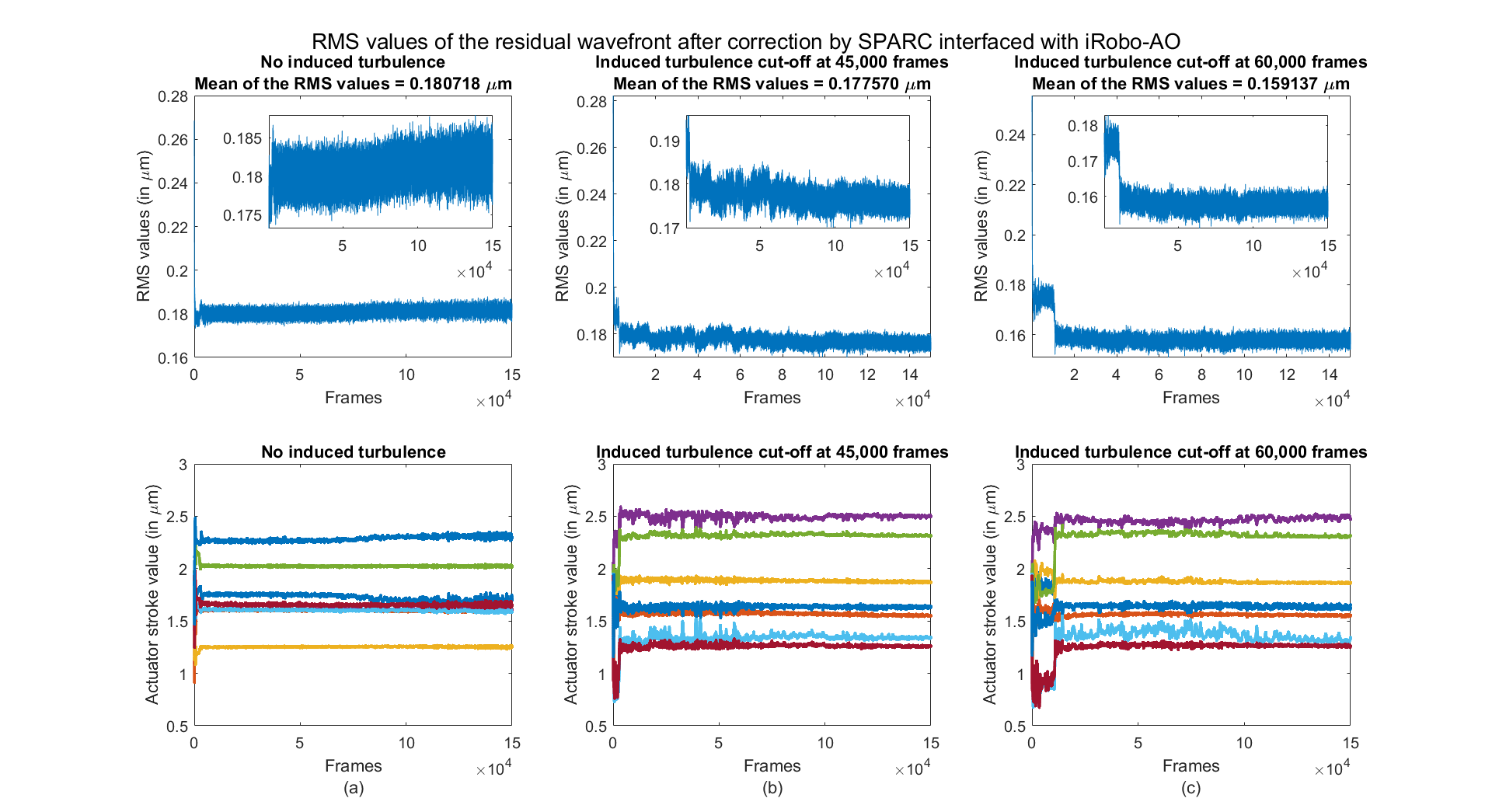}
\caption{RMS values of the residual wavefront at the WFS {(with the insets showing the same from the 500th frame onwards after the residual RMS values have stabilized)}, and the actuator stroke values of the central row of actuators (actuator numbers 61-68). Time series of observations in the lab a) when no turbulence was induced b) when the power supply to the resistance was switched off at 45,000 frames c) when the power supply to the resistance was switched off at 60,000 frames.}
\label{fig:iroboao_times}
\end{figure}

Fig~\ref{fig:iroboao_times} shows the results of the residual WFE and the actuator stroke values from a set of observations of SPARC interfaced with iRobo-AO. {The insets in the top row of figures show the RMS values between from the 500th to the last frame, after the AO control loop has stabilized the residual RMS values over the initial 100-200 frames.} Without any induced turbulence under the laboratory environment, the standard deviation in the RMS WFE is around 1.7 nm (in Fig~\ref{fig:iroboao_times}a). A 10 W resistor is used to induce turbulence in the optical path by generating a temperature inhomogeneity during the observation run. After the resistor is powered on, it is allowed enough time to reach a steady state temperature before the observations start. The resistor is switched off after a finite time to verify whether the reduction in the turbulence causes a reduction in the variation in DM actuator stroke and RMS WFE of the residual wavefront. The mean of the RMS values of the wavefront ranged from roughly 0.16 $\mu$m to 0.18 $\mu$m as shown in Fig~\ref{fig:iroboao_times}. The on-sky RMS values of the Robo-AO system at Palomar varies from 0.137 $\mu$m to 0.241 $\mu$m\cite{baranec2012}, and the mean of the RMS value from measurements with the iRobo-AO laboratory platform using the native software was around 0.13 $\mu$m, without any induced turbulence. The standard deviation of the RMS values of the wavefront is computed (while the control loop is running) to estimate the ability of the control loop to stabilize the wavefront. The standard deviation of the RMS values when the resistor was switched off after 45,000 frames or about 80 seconds into the observation (in Fig~\ref{fig:iroboao_times}b) is about 2.3 nm (1.2\% of the mean of the RMS value). When the resistor was switched off after 60,000 frames or about 110 seconds into the observation (in Fig~\ref{fig:iroboao_times}c), the standard deviation in the RMS WFE is about 4.4 nm (2.8\% of the mean of the RMS value).

\section{Summary}
\label{sec:disc}

SPARC is a pathfinder for FPGA-based real-time kernels for realistic AO systems that will be needed by the next generation of extremely large telescopes, which will require MCAO and Extreme-AO. The proof of concept implementation uses a cheap commercial FPGA development board to perform a full AO reconstruction for a $2601\times2500$ matrix (for $50\times50$ subapertures) in a median time of 1.283 ms (out of which 1 ms is taken for the retrieval of the reconstruction matrix from the external DDR3 memory). By using our platform on already available hardware which uses faster FPGAs and memories, it can meet the requirements of challenging AO implementations. {FPGAs can compete with GPUs in terms of memory bandwidth and computational performance.}

The system is designed to be modular so that the slope computation algorithm can be replaced by alternative methods (matched filtering, weighted CoG etc.), while preserving the parallel computational capabilities of the WPU. We have tested the system upto $50\times50$ subapertures for a single channel WFS, but the system is scalable for any subaperture size and any DDR type of memory. The system is limited only by the memory bandwidth and the logic resources available on the FPGA. The advent of serial memories like the high bandwidth memory or HBM (explained in Part 2\cite{SPARC2}) increases the memory bandwidth limits of today's FPGAs by a factor of 16-18 times, compared to our development board. The Xilinx VU35P or the VU37P (commercially available today) will not need an external memory to store the reconstruction matrix, and will be able to provide a theoretical memory bandwidth of 460 GB/s\cite{xilinx2017}. With the proportional increase in logic resources in these Xilinx chips, they can be expected to perform the AO reconstruction for a $50\times50$ subaperture frame in less than 100 $\mu$s. {GPUs and CPUs will still be better for conventional AO implementations on large telescopes in the near future, but for applications which demand a predictable latency (like Extreme-AO), the HBM-FPGA implementation will provide a better solution. We will be perfecting the platform to have the capability to actually address the computational needs of a thirty meter class telescope and have a shorter development cycle compared to that of conventional AO kernels, through the future versions of SPARC. The future version of SPARC is also planned to have the interface to directly connect to different WFS camera interfaces (like IEEE1394b/Firewire, CameraLink, USB 3.0, 10GbE or even through a PCIe backplane).}

\appendix    % this command starts appendixes

\acknowledgments 
We are thankful to the Department of Science and Technology, University Grants Commission and the Infosys Foundation for funding and supporting this project.
%%%%% References %%%%%

%\nocite{*}
\bibliography{SPARC_1}   % bibliography data in report.bib

\begin{thebibliography}{10}

\bibitem{herriot2014}
G.~Herriot, D.~Andersen, J.~Atwood, {\em et~al.}, ``{NFIRAOS: first facility AO
  system for the Thirty Meter Telescope},''  (2014).

\bibitem{diolati2008}
E.~Diolaiti, J.-M. Conan, I.~Foppiani, {\em et~al.}, ``{A preliminary overview
  of the multiconjugate adaptive optics module for the E-ELT},''  (2008).

\bibitem{tokovinin2004}
A.~Tokovinin, ``{Seeing Improvement with Ground-Layer Adaptive Optics},'' {\em
  Publications of the Astronomical Society of the Pacific} {\bf 116}(824), 941
  (2004).

\bibitem{marcos2014}
B.~A.~M. Marcos A.~van Dam, Antonin H.~Bouchez, ``{Wide field adaptive optics
  correction for the GMT using natural guide stars},''  (2014).

\bibitem{hinz2010}
P.~M. Hinz, A.~Bouchez, M.~Johns, {\em et~al.}, ``{The GMT adaptive optics
  system},'' in {\em proc. SPIE},   {\bf 7736}, 77360C  (2010).

\bibitem{chin2014}
J.~C. Chin, P.~Wizinowich, E.~Wetherell, {\em et~al.}, ``{Laser guide star
  facility developments at WM Keck Observatory},'' in {\em SPIE Astronomical
  Telescopes+ Instrumentation},  914808--914808, International Society for
  Optics and Photonics  (2014).

\bibitem{wizinowich2008}
P.~Wizinowich, R.~Dekany, D.~Gavel, {\em et~al.}, ``{WM Keck Observatory's
  next-generation adaptive optics facility},'' in {\em Proc. SPIE},   {\bf
  7015}, 701511  (2008).

\bibitem{max1997}
C.~E. Max, S.~S. Olivier, H.~W. Friedman, {\em et~al.}, ``Image improvement
  from a sodium-layer laser guide star adaptive optics system,'' {\em Science}
  {\bf 277}(5332), 1649--1652  (1997).

\bibitem{baranec2012}
C.~Baranec, R.~Riddle, A.~Ramaprakash, {\em et~al.}, ``{Robo-AO: autonomous and
  replicable laser-adaptive-optics and science system},'' in {\em SPIE
  Astronomical Telescopes+ Instrumentation},  844704--844704, International
  Society for Optics and Photonics  (2012).

\bibitem{jensen2017}
R.~Jensen-Clem, D.~A. Duev, R.~Riddle, {\em et~al.}, ``{The Performance of the
  Robo-AO Laser Guide Star Adaptive Optics System at the Kitt Peak 2.1-m
  Telescope},'' {\em arXiv preprint arXiv:1703.08867}   (2017).

\bibitem{wizinowich2006}
P.~L. Wizinowich, D.~Le~Mignant, A.~H. Bouchez, {\em et~al.}, ``{The WM Keck
  Observatory laser guide star adaptive optics system: overview},'' {\em
  Publications of the Astronomical Society of the Pacific} {\bf 118}(840), 297
  (2006).

\bibitem{rousset2000}
G.~Rousset, F.~Lacombe, P.~Puget, {\em et~al.}, ``{Status of the VLT Nasmyth
  adaptive optics system (NAOS)},'' in {\em Astronomical Telescopes and
  Instrumentation},  72--81, International Society for Optics and Photonics
  (2000).

\bibitem{dekany2013}
R.~Dekany, J.~Roberts, R.~Burruss, {\em et~al.}, ``{PALM-3000: exoplanet
  adaptive optics for the 5 m Hale telescope},'' {\em The Astrophysical
  Journal} {\bf 776}(2), 130  (2013).

\bibitem{rigaut2013}
F.~Rigaut, B.~Neichel, M.~Boccas, {\em et~al.}, ``{Gemini multiconjugate
  adaptive optics system review--I. Design, trade-offs and integration},'' {\em
  Monthly Notices of the Royal Astronomical Society} {\bf 437}(3), 2361--2375
  (2013).

\bibitem{beuzit2008}
J.-L. Beuzit, M.~Feldt, K.~Dohlen, {\em et~al.}, ``{SPHERE: a planet finder
  instrument for the VLT},'' in {\em proc. SPIE},   {\bf 7014}, 701418  (2008).

\bibitem{gavel2002lick}
D.~T. Gavel, E.~Gates, C.~Max, {\em et~al.}, ``{Recent science and engineering
  results with the laser guidestar adaptive optics system at Lick
  Observatory},'' tech. rep., Lawrence Livermore National Lab., CA (US)
  (2002).

\bibitem{morzinski2014}
K.~M. Morzinski, L.~M. Close, J.~R. Males, {\em et~al.}, ``{MagAO: Status and
  on-sky performance of the Magellan adaptive optics system},'' in {\em SPIE
  Astronomical Telescopes+ Instrumentation},  914804--914804, International
  Society for Optics and Photonics  (2014).

\bibitem{rigaut1998}
F.~Rigaut, D.~Salmon, R.~Arsenault, {\em et~al.}, ``{Performance of the
  Canada-France-Hawaii Telescope Adaptive Optics Bonnette},'' {\em Publications
  of the Astronomical Society of the Pacific} {\bf 110}(744), 152  (1998).

\bibitem{minowa2010}
Y.~Minowa, Y.~Hayano, S.~Oya, {\em et~al.}, ``{Performance of Subaru adaptive
  optics system AO188},'' in {\em proc. SPIE},   {\bf 7736}, 77363N  (2010).

\bibitem{ellerbroek2013}
B.~{Ellerbroek}, ``{Adaptive Optics for the Thirty Meter Telescope},'' in {\em
  Proceedings of the Third AO4ELT Conference},  S.~{Esposito} and L.~{Fini},
  Eds., 21  (2013).

\bibitem{wang2013}
L.~{Wang}, ``{Design and Testing of GPU based RTC for TMT NFIRAOS},'' in {\em
  Proceedings of the Third AO4ELT Conference},  S.~{Esposito} and L.~{Fini},
  Eds., 17  (2013).

\bibitem{SPARC2}
A.~Surendran, M.~P. Burse, A.~N. Ramaprakash, {\em et~al.}, ``{Scalable
  Platform for Adaptive optics Real-time Control (SPARC) Part 2: Field
  Programmable Gate Array (FPGA) implementation and performance},''  (2018).
\newblock Under review.

\bibitem{gavel2002}
D.~T. Gavel, ``Adaptive optics control strategies for extremely large
  telescopes,''  (2002).

\bibitem{sevin2014}
A.~Sevin, D.~Perret, D.~Gratadour, {\em et~al.}, ``{Enabling technologies for
  GPU driven adaptive optics real-time control},'' in {\em Proc. SPIE},   {\bf
  9148}, 91482G  (2014).

\bibitem{asurendran2015}
A.~Surendran, M.~P. Burse, A.~Ramaprakash, {\em et~al.}, ``Development of a
  scalable generic platform for adaptive optics real time control,'' in {\em
  International Conference on Optics \& Photonics 2015},  965425--965425,
  International Society for Optics and Photonics  (2015).

\bibitem{nvidia2014}
Nvidia, ``{Tesla K10 GPU accelerator},'' tech. rep., Nvidia  (2014).

\bibitem{sedmak2004}
G.~Sedmak, ``{Implementation of fast-Fourier-transform-based simulations of
  extra-large atmospheric phase and scintillation screens},'' {\em Applied
  optics} {\bf 43}(23), 4527--4538  (2004).

\bibitem{herrmann1980}
J.~Herrmann, ``Least-squares wave front errors of minimum norm,'' {\em JOSA}
  {\bf 70}(1), 28--35  (1980).

\bibitem{roddier1999}
F.~Roddier, {\em Adaptive optics in astronomy}, Cambridge university press
  (1999).

\bibitem{altera2013}
Altera, ``{Radar Processing: FPGAs or GPUs?},'' tech. rep.  (2013).

\bibitem{xilinx2017}
Xilinx, ``{UltraScale Architecture and Product Data Sheet: Overview},'' tech.
  rep.  (2017).

\end{thebibliography}
\bibliographystyle{spiejour}   % makes bibtex use spiejour.bst

%%%%% Biographies of authors %%%%%

\listoffigures
\listoftables

\end{spacing}
\end{document}